# Exploring the Accuracy of MIRT Scale Linking Procedures for Mixed-format Tests

Zhen Li, Texas Education Agency

Haiqin Chen, American Dental Association

Tianli Li, ACT Inc.

## Abstract

This study investigates the accuracy of Stocking-Lord scale linking procedures for UIRT and MIRT models with common-item nonequivalent-group design for mixed-format tests under two anchor conditions: MC-Only and MC-CR across three different levels of format effects (FEs). Results provide recommendations on the appropriateness of UIRT and two MIRT models when FEs present under each anchor scenario.

## Purpose

Mixed-format tests are widely used in K-12 testing. However, the multiple-choice (MC) and constructed-response (CR) subtests of the same content may measure different constructs which can lead to format effects (FEs) at test level (Traub, 1993). FEs may introduce equating errors when unidimensional IRT (UIRT) models are used; and/or when MC-only anchor items are selected. Previous studies suggested that multidimensional IRT (MIRT) models tend to perform better when tests are more multidimensional (e.g., Yao & Boughton, 2009). Other research also underlined that anchor items should represent not only the content and statistical specifications, but also the dimensional structure, suggesting that both MC and CR anchor items should be used (Lee & Brossman, 2012; Lee & Lee, 2016) for mixed-format tests.

Stocking-Lord (SL) method is one of most commonly used and recommended UIRT linking procedures. Yao & Boughton (2009) proposed a test response function (TRF) matching procedure, a multidimensional extension of the Stocking-Lord method (M-SL) for mixed-format tests. Their study investigated MC-CR and MC-only anchor item scenarios across multiple multidimensional population distributions. They found smaller estimation errors when (1) the correlation between dimensions are lower (e.g., .0 or .5); (2) using MC-CR anchor items or MC-only anchor items with score points that are close to the total score points of MC-CR anchor items.

The purpose of this simulation study was to determine:

1) Whether using MC-CR anchor items increases the parameter recovery accuracy;

2) Whether MIRT models perform better than UIRT models when format effects present;

3) Which MIRT model produces better recovery: simple-structure or bifactor?

## Methods

### *Data Generation*

To be neutral to the UIRT or MIRT models, a simple structure MIRT model was used to generate response data (Kolen, Wang, & Lee, 2012; Lee & Brossman, 2012) using real parameter estimates from separate calibrations of MC and CR subtests of a mathematics mixed-format test with 3PL and generalized partial credit (GPC) models. The latent traits representing MC and CR dimensions were simulated from bivariate standard normal distributions. Each simulation condition was replicated 20 times with 3000 examinees in each run. Table 1 shows the item parameter estimates used in simulation.

[Insert Table 1 Here]

### *Factors Investigated*

Two anchor-sets were investigated: (1) MC-only; (2) MC-CR.

FEs were manipulated by varying the correlations ($\rho$) between MC and CR format-specific factors. Three levels of correlations between the MC and CR factors were considered: 0.5, 0.8, and 1.0, representing large, medium, and no FEs.

Two MIRT models were investigated: (1) Bi-factor model (Cai et al, 2011) allows items load on both general and two format-specific factors and assumes that all the factors are mutually orthogonal. (2) Simple structure model has two format-specific factors with MC items load on MC factor only and CR items load on CR factor only. Factors are allowed to be correlated.

The multidimensional 3PL (M-3PL; Reckase, 2009) and partial credit models (M-2PP; Yao & Schwarz, 2006), bi-factor 3PL and GPC models (Cai et al., 2011) were used to obtain the estimates for MC and CR items from the two MIRT models.

To link the item and ability parameters, the TRF matching method was used to determine the estimates of the D by D rotation matrix A and D-element location vector $\vec{B}$ by minimizing the difference between TRFs. Then the parameter estimates of the new form can be transformed using the formulas in Yao & Boughton (2009).

The computer programs BMIRT (Yao, 2003) and LinkMIRT (Yao, 2004) were used to obtain the estimates of parameter and elements of A and $\vec{B}$.

Rackase (2009) indicated that the accuracy of the transformation of the item parameters from the new test calibration to the base coordinate system can be evaluated by comparing the common item parameter estimates after transformation with the values from the calibration in the base coordinate system. In this study, the rotation matrix and the location vector were examined first. Then average root mean square differences (ARMSDs) of the common item parameters were checked to examine the accuracy of the transformations. In addition, Population mean and variance-covariance recovery for simple-structure MIRT model was used to examine the performance of multidimensional extension of SL method under MC-CR and MC-only anchor-sets.

## Results

Table 2 provides the equating constants, A and B estimated using UIRT SL method across three levels of FEs and two anchor item conditions. When there was no or low FEs, $A$s from both anchor-item conditions were close to one while $B$s were close to zero. As FEs increase (e.g., $\rho$=0.5), $A$ from MC-only anchor condition was under-estimated, and $B$ was over-estimated. Both $A$ and $B$ from MC-CR anchor-set were not affected.

Table 3 presents the average root mean square differences (ARMSDs) for anchor items under two anchor scenarios and across three levels of FEs. When there was no or low FEs, the ARMSDs for MC $a$, $b$-parameter estimates were similar under two anchor conditions, while the values under MC-CR condition were slightly lower. As FEs increase (e.g., $\rho$=0.5), ARMSDs are clearly increased. In all cases, FEs have a bigger effect on $a$-parameter than $b$-parameter estimates. The ARMSDs were smaller under MC-CR conditions than those under the MC-only conditions. For CR item parameter estimates, the FEs have a big effect on $a$ and step parameter estimates, particularly when FEs are large (e.g., $\rho$=0.5). Note that under MC-only anchor condition, there are no CR anchor items.

[Insert Tables 2 and 3 Here]

Table 4 shows the MIRT transformation matrix (A) and location vector ($\vec{B}$) under two anchor scenarios across three levels of FEs for bifactor and simple-structure MIRT models.

For bifactor MIRT model, FEs do not have a big effect on the estimations of A and $\vec{B}$. Using MC-CR anchor items, all the entries of A and $\vec{B}$ were similar across three levels of FEs, which were close to identity matrix and 0 vector. With MC-only anchor items, some entries of A and $\vec{B}$ were over-estimated when no FEs presents. The entries in A and $\vec{B}$ under MC-only were in general larger (over-estimated) than those under MC-CR conditions, indicating MC-only anchor-items resulted in less accurate estimation of A and $\vec{B}$.

For simple-structure MIRT model, FEs do not have a big effect on the estimates of A and $\vec{B}$. Under two anchor conditions, all entries were similar across three FEs levels. The second entry of $\vec{B}$ was over-estimated across three levels of FEs under MC-only condition.

In both bifactor and simple-structure cases, using MC-CR anchor-item produces more accurate estimations of A and $\vec{B}$. For bifactor model, when MC-only anchor-item was used, the A and $\vec{B}$ were slightly over-estimated. For simple-structure model, the second entries of $\vec{B}$ were over-estimated across all levels of FEs.

[Insert Table 4 Here]

Table 5 provides the average root mean square differences (ARMSDs) between parameter estimates from the common items after transformation and the values from the calibration in the base coordinate system for both bifactor and simple-structure MIRT models.

For bifactor MIRT model, for MC item parameter transformation, even though there was no big difference on transformation matrices and location vectors between MC-CR and MC-only anchor-sets, there were bigger ARMSDs associate with larger FEs. For CR item $a1$ and $a2$, when MC-CR anchor-set was used, bigger ARMSDs were associated with small FE. For $t1$ and $t2$, the patterns were inconsistent but generally the differences between levels of FEs were small.

For simple-structure MIRT model, for MC item parameter transformation, under both MC-CR and MC-only cases, for $a$ parameter, Fes had no effects. However, bigger ARMSDs associated with no FEs for $d$ were found. For CR item parameter transformation, when MC-CR anchor was used, FEs had no effect on $a$ and $d$.

[Insert Table 5 Here]

Table 6 shows the recovery of the population mean and variance-covariance matrix from simple-structure MIRT model using multidimensional extension of SL scale transformation method under three FE levels and two anchor scenarios. Overall, the mean vectors were well recovered across three levels of FEs; all were close to a zero vector when an MC-CR anchor-set was used. However, when MC-only anchor-set was used, one second element of the mean vector was over-estimated. For variance-covariance matrix, for both anchor scenarios, the values were more inflated when no or small FEs present. The values under MC-only anchor scenario were more inflated than those from MC-CR anchor scenario when no FE presents.

[Insert Table 6 Here]

## Conclusion

Anchor format was more important when UIRT scale linking method was used, particularly when large FE presents. In this case, using MC-CR anchor-set will produce more accurate transformation.

MIRT linking method performed better than UIRT linking method when multidimensionality presents. The format effects on transformation matrix A and location vector B were small. However, FEs do affect scaling accuracy slightly through parameter estimation. In addition, FEs affect different parameter transformation differently. The amount of ARMSD from two MIRT models were comparable.

Under certain conditions, using MC-CR anchor-set does result in more accurate transformation matrix and location vector estimates for both MIRT models. For both MIRT models, under both anchor format conditions, the ARMSDs were similar for three levels of multidimensionality.

Item parameters that have extreme values tend to have larger RMSDs. Mean and variance-covariance matrix from simple-structure MIRT model also recovered better when MC-CR anchor set was used.

## Contribution

This study provides information to practitioners and researchers on the appropriateness of the UIRT or MIRT scale linking methods when varying levels of FEs are present under two anchor-set scenarios.

Table 1

*Statistics of Item Parameters for New and Old Forms*

| | *n* | *Mean* | *SD* | *Min* | *Max* | *Mean* | *SD* | *Min* | *Max* |
|---|---|---|---|---|---|---|---|---|---|
| MC items | | New form | | | | Old form | | | |
| a | 40 | 0.998 | 0.313 | 0.417 | 1.581 | 0.992 | 0.285 | 0.418 | 1.581 |
| b | 40 | 0.203 | 0.780 | -2.313 | 1.777 | 0.065 | 0.773 | -2.363 | 2.050 |
| c | 40 | 0.238 | 0.090 | 0.066 | 0.422 | 0.226 | 0.079 | 0.062 | 0.380 |
| CR items | | New form | | | | Old form | | | |
| a | 12 | 1.214 | 0.331 | 0.748 | 1.896 | 1.264 | 0.605 | 0.263 | 2.240 |
| $\boldsymbol{\delta_{j1}}$[a] | 12 | -0.074 | 0.383 | -0.640 | 0.635 | 0.253 | 0.995 | -0.922 | 2.735 |
| $\boldsymbol{\delta_{j2}}$[a] | 12 | -0.302 | 0.798 | -2.045 | 0.806 | -0.517 | 1.020 | -2.656 | 0.806 |

[a]$\delta_{jk}$ refers to transition location parameter between the $k$th category and the $(k-1)th$ category under the generalized partial credit (GPC) model.

Table 2

*Scaling Constants, A and B from UIRT model under Two Anchor Scenarios across Three Levels of FEs*

| | SL_A | SL_B |
|---|---|---|
| | FE=1.0 | |
| MC-only | 1.00 | -0.01 |
| MC-CR | 1.01 | 0.01 |
| | FE=0.8 | |
| MC-only | 0.99 | 0.00 |
| MC-CR | 1.00 | 0.01 |
| | FE=0.5 | |
| MC-only | 0.95 | 0.05 |
| MC-CR | 1.00 | 0.01 |

Table 3

*ARMSDs for Anchor-Item Parameter Estimates under Two Anchor Scenarios across Three Levels of FEs*

| | FE=0.5 | FE=0.8 | FE=1.0 |
|---|---|---|---|
| | MC-a | | |
| MC-CR | 0.24 | 0.12 | 0.11 |
| MC-only | 0.26 | 0.13 | 0.12 |
| | MC-b | | |
| MC-CR | 0.18 | 0.10 | 0.09 |
| MC-only | 0.19 | 0.11 | 0.10 |
| | CR-a | | |
| MC-CR | 0.99 | 0.76 | 0.59 |
| | CR-b | | |
| MC-CR | 0.11 | 0.05 | 0.05 |
| | CR-step | | |
| MC-CR | 0.44 | 0.13 | 0.05 |

*Note.* The CR items were only examined under MC-CR anchor scenario since there are no common CR items in MC-Only case.

Table 4(a)

*Transformation Matrix A and Location Vector $\vec{B}$ from the Bi-factor MIRT Model under Two Anchor Scenarios across Three Levels of FEs*

| Bifactor | Transformation Matrix A | | | | | | | | | Location Vector $\vec{B}$ | | |
|---|---|---|---|---|---|---|---|---|---|---|---|---|
| MC-CR anchor | | | | | | | | | | | | |
| FE (ρ) | a11 | a12 | a13 | a21 | a22 | a23 | a31 | a32 | a33 | b1 | b2 | b3 |
| 0.5 | 1.0 | 0.0 | 0.1 | 0.1 | 1.1 | 0.2 | 0.1 | 0.2 | 1.0 | -0.1 | 0.2 | 0.1 |
| 0.8 | 1.0 | 0.0 | 0.0 | 0.1 | 1.1 | 0.2 | 0.1 | 0.1 | 1.0 | -0.1 | 0.1 | 0.0 |
| 1.0 | 1.0 | 0.0 | 0.0 | 0.1 | 1.1 | 0.1 | 0.1 | 0.1 | 1.0 | 0.0 | 0.1 | 0.1 |
| MC-only anchor | | | | | | | | | | | | |
| 0.5 | 1.0 | 0.0 | 0.2 | 0.0 | 1.1 | 0.2 | 0.2 | 0.2 | 1.1 | -0.1 | 0.1 | 0.1 |
| 0.8 | 1.0 | 0.0 | 0.2 | 0.1 | 1.1 | 0.2 | 0.2 | 0.2 | 1.1 | -0.1 | 0.2 | 0.1 |
| **1.0** | **1.1** | **0.0** | **0.3** | **0.2** | **1.2** | **0.2** | **0.2** | **0.2** | **1.1** | **0.0** | **0.2** | **0.2** |

*Note.* The MC, CR, and General factors are assumed to be uncorrelated in bi-factor MIRT model.

Table 4(b)

*Transformation Matrix A and Location Vector $\vec{B}$ from the Simple Structure MIRT under Two Anchor Scenarios across Three Levels of FEs*

| Simple-Structure | Transformation Matrix A | | | | Location Vector $\vec{B}$ | |
|---|---|---|---|---|---|---|
| MC-CR anchor | | | | | | |
| FE (ρ) | a11 | a12 | a21 | a22 | b1 | b2 |
| 0.5 | 1.0 | 0.2 | 0.2 | 1.1 | 0.0 | 0.0 |
| 0.8 | 1.1 | 0.2 | 0.2 | 1.0 | 0.0 | 0.0 |
| 1.0 | 1.0 | 0.1 | 0.1 | 1.0 | 0.0 | 0.0 |
| MC-only anchor | | | | | | |
| 0.5 | 1.0 | 0.2 | 0.2 | 1.1 | 0.0 | **0.2** |
| 0.8 | 1.0 | 0.2 | 0.2 | 1.1 | 0.0 | **0.2** |
| 1.0 | 1.1 | 0.2 | 0.2 | 1.1 | 0.0 | **0.2** |

*Note.* The MC and CR factors can be correlated in simple structure MIRT model.

Table 5(a)

*ARMSDs Between Parameter Estimates from the Common Items after Transformation with the Values from the Calibration in the Base Coordinate System under the Two Anchor scenarios and across Three Levels of FE for Bifactor MIRT Models*

| Bifactor MIRT model | | | | | | | | | | |
|---|---|---|---|---|---|---|---|---|---|---|
| | MC Item Parameters | | | | | | CR Item Parameters | | | |
| | a1 | | a2 | | d | | a1 | a2 | t1 | t2 |
| Rho | MC-CR | MC-only | MC-CR | MC-only | MC-CR | MC-only | MC-CR | MC-CR | MC-CR | MC-CR |
| 0.5 | 0.23 | 0.21 | 0.21 | 0.21 | 0.17 | 0.17 | 0.06 | 0.08 | 0.09 | 0.10 |
| 0.8 | 0.19 | 0.18 | 0.19 | 0.19 | 0.15 | 0.14 | 0.08 | 0.08 | 0.11 | 0.09 |
| 1.0 | 0.18 | 0.18 | 0.16 | 0.16 | 0.14 | 0.14 | 0.07 | 0.09 | 0.09 | 0.09 |

*Note.* The CR items were only examined under MC-CR anchor scenario since there are no common CR items in MC-Only case.

Table 5(b)

*ARMSDs Between Parameter Estimates from the Common Items after Transformation with the Values from the Calibration in the Base Coordinate System under the Two Anchor scenarios and across Three Levels of FE for Simple-Structure MIRT Models*

| Simple-Structure MIRT model | | | | | | |
|---|---|---|---|---|---|---|
| | MC Item Parameters | | | | CR Item Parameters | |
| | a | | d | | a | t |
| Rho | MC-CR | MC-only | MC-CR | MC-only | MC-CR | MC-CR |
| 0.5 | 0.18 | 0.18 | 0.13 | 0.13 | 0.06 | 0.09 |
| 0.8 | 0.17 | 0.17 | 0.13 | 0.13 | 0.08 | 0.11 |
| 1.0 | 0.18 | 0.18 | 0.14 | 0.14 | 0.07 | 0.09 |

*Note.* The CR items were only examined under MC-CR anchor scenario since there are no common CR items in MC-Only case.

Table 6

*Transformed Population Mean and Var-Covariance Matrix under Simple-Structure MIRT Model*

| MC-CR | | | | | | |
|---|---|---|---|---|---|---|
| Rho | Mean | | Variance-covariance | | | |
| 0.5 | 0.0 | 0.0 | 1.3 | 0.9 | 0.9 | 1.5 |
| 0.8 | 0.0 | 0.0 | 1.7 | 1.3 | 1.3 | 1.4 |
| 1.0 | 0.0 | 0.0 | 1.3 | 1.2 | 1.2 | 1.3 |
| MC-only | | | | | | |
| | Mean | | Variance-covariance | | | |
| 0.5 | 0.0 | 0.2 | 1.3 | 0.9 | 0.9 | 1.5 |
| 0.8 | 0.0 | 0.2 | 1.4 | 1.3 | 1.3 | 1.7 |
| 1.0 | 0.0 | 0.2 | 1.8 | 1.6 | 1.6 | 1.8 |